\documentclass[final,number,times, sort&compress]{elsarticle}

\usepackage{hyperref}
\usepackage{amsmath}
\usepackage{amssymb}
\journal{Physica B: Condensed Matter}

\hypersetup{colorlinks=true, breaklinks=true, linkcolor=blue, menucolor=blue, urlcolor=blue}

\begin{document}

\begin{frontmatter}

\title{Novel Neutron Detectors based on the Time Projection Method}

\author[bo,us]{M. K\"ohli\corref{cor1}}
\ead{koehli@physik.uni-bonn.de}
\author[bo]{K. Desch}
\author[bo]{M. Gruber}
\author[bo]{J. Kaminski}
\author[bo]{F.P. Schmidt}
\author[bo]{T. Wagner}

\address[bo]{Physikalisches Institut, University of Bonn, Nussallee 12, 53115 Bonn, Germany}
\address[us]{Physikalisches Institut, Heidelberg University, Im Neuenheimer Feld 226, 69120 Heidelberg, Germany}

\cortext[cor1]{Corresponding author}

\begin{abstract}
We present the first prototype of a novel thermal neutron detector using the time projection method. The system consists of 8~TimePix ASICS with postprocessed InGrid meshes. Each ASIC has $256 \times 256$~pixels of 55\,$\muup$m$\times$55\,$\muup$m in size with the capability to measure charge or time. This allows to visualize entire conversion particle tracks with their spatial and time information and, by using event reconstruction algorithms, discriminate against the background of others. By using the Scalable Readout System the detector as presented here could also be upscaled to much larger active areas. In the current configuration we could  achieve a spatial resolution of $\sigma = (115\pm8)\,\muup\text{m}$.

\end{abstract}


\end{frontmatter}


\section{Introduction}
The world of detectors used in thermal neutron scattering instrumentation has changed. Much of what once was established has been discarded. For them now substitutional technologies have been presented. It began with production of tritium and peaked at the crisis of helium-3. Part of it was given to sciences for basic or applied research. Part for the industry, explorating oil deep in the rocks. And the largest part was given to homeland security, which above all else demanded for it for the protection against hazards. With the stockpile nearly exhausted, alerts~\cite{Fed09} on the future helium-3 supply\cite{He3crisis}, critical to perspectives of large-scale research infrastructures~\cite{TdrESS}, started the run on substitutional technologies~\cite{ZeitelhackAlternatives}. Most of the solutions~\cite{NeutronESS} could be adapted from developments of particle physics and are comprised of one or more layers of Boron-10~\cite{Piscitelli2015}. The Time Projection Method~\cite{yTPC} achieves highest spatial resolution along with time information by high granular track reconstruction of the ions released from solid Boron-10 or other types of conversion products~\cite{yTPC2}. Based on the experience of the CASCADE Detector~\cite{Cascade} installed at the Spin-Echo~\cite{Nrse1987} instruments RESEDA~\cite{ResedaJLSRF} and MIRA~\cite{MiraJLSRF} at FRM II in Germany, the University of Bonn now develops a novel system using Time Projection Chamber~\cite{timeprojectionchamber} technology with a pixelized readout~\cite{pixelTPCIEEE}. This detector based on TimePix~\cite{TimepixLlopart} chips aims for sub-100 micrometer resolution at medium count rates and increasing of the efficiency of a single layer of Boron-10 by stacking several consecutive units in beam direction.

\section{Neutron detection by solid converters}

Thermal neutrons ($\lambda = 1.8\, \text{\AA },\ E_{\mathrm{kin}}= 25.2\,\mathrm{meV}$) are in equilibrium with their environment and so the change of entropy in the sensitive active medium required for a significant direct detection is too small. 
Converters are used to obtain a clear and strong signature by processes which are specifically sensitive to thermal neutrons such as nuclear absorption followed by immediate de-excitation. Only a few isotopes exhibit cross sections reasonably high and products of charged particles unambiguous enough such as $^3$He, $^6$Li, $^{10}$B. 
\begin{figure}[ht]
\centering 
\includegraphics[width=0.95\linewidth]{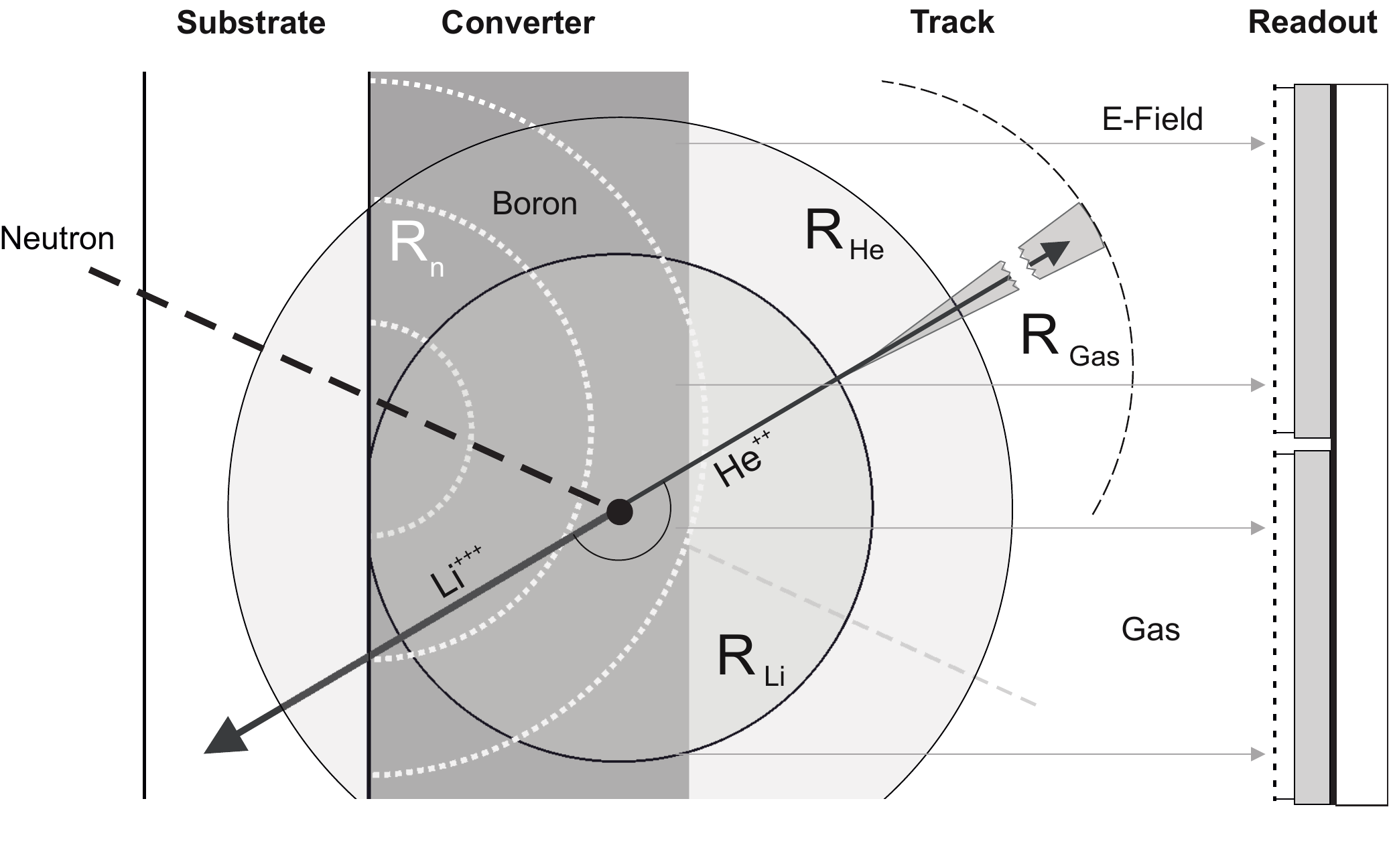}
\caption{Hybrid boron converter layer system (not to scale): a neutron enters at an angle $\theta$ normal to the surface. The absorption probability scales wavelength dependent by the Beer-Lambert law. After absorption in boron lithium (Li) and helium (He) ions with specific energies are created. In the converter medium they lose energy by collisions leading to a Bragg distributed range, different for both agents. After leaving the solid layer an ionization track is produced according to the remaining energy at the boundary. The primary ionization is projected onto a readout structure with a gas amplification stage in front.}
\label{fig:schematicBoron}
\end{figure}
In contrast to gaseous helium, solids like boron and lithium require a hybrid detector with conversion and amplification stages separated. The particles emitted are charged fragments of high energy loss by dense ionization, thus converter layers have to be as thin as micrometers. Meanwhile these systems are understood and optimized by means of analytical modeling~\cite{McGregor2003272Design} or Monte Carlo transport simulations~\cite{thermalNeutronMCNP}, detailed descriptions for specific geometries can also be found in~\cite{JalousieEfficiency} and \cite{b10filmCalc}.
For now a simple single layer as shown in Fig.~\ref{fig:schematicBoron} is considered. The capture cross section of boron at thermal energies is $(3844\pm3)$\,barn~\cite{endfRef} and scales linearly by wavelength $\lambda$. Two absorption reactions occur:%
\begin{align*}
& ^{10}\mathrm{B} + \mathrm{n} \rightarrow & \\
& ^7\mathrm{Li}(0.84\,\mathrm{MeV}) + \alpha(1.472\,\mathrm{MeV}) + \gamma (0.48\,\mathrm{MeV}) &(93.6\,\%),\\
&	 ^7\mathrm{Li}(1.013\,\mathrm{MeV}) + \alpha(1.776\,\mathrm{MeV}) &(6.4\,\%)		.
\end{align*}
In the boron layer itself the isotropically and back-to-back emitted particles lose their kinetic energy mainly by ionization. The average maximum ranges are from (1.69-1.90)\,$\muup$m for lithium to (3.27-4.05)\,$\muup$m for helium ions with a straggling variation in the order of $\sigma \approx 5$\,\% according to range calculations by SRIM~\cite{SRIM}. These values, normalized to bulk density, are similar for boron carbide and boron and limit the active volume of each layer to spheres, which are depicted in Fig.~\ref{fig:schematicBoron}. In the case of multi-layer detectors the amount of inactive converter material decreases the efficiency.  
\begin{figure}[ht]
\centering 
\includegraphics[width=\linewidth]{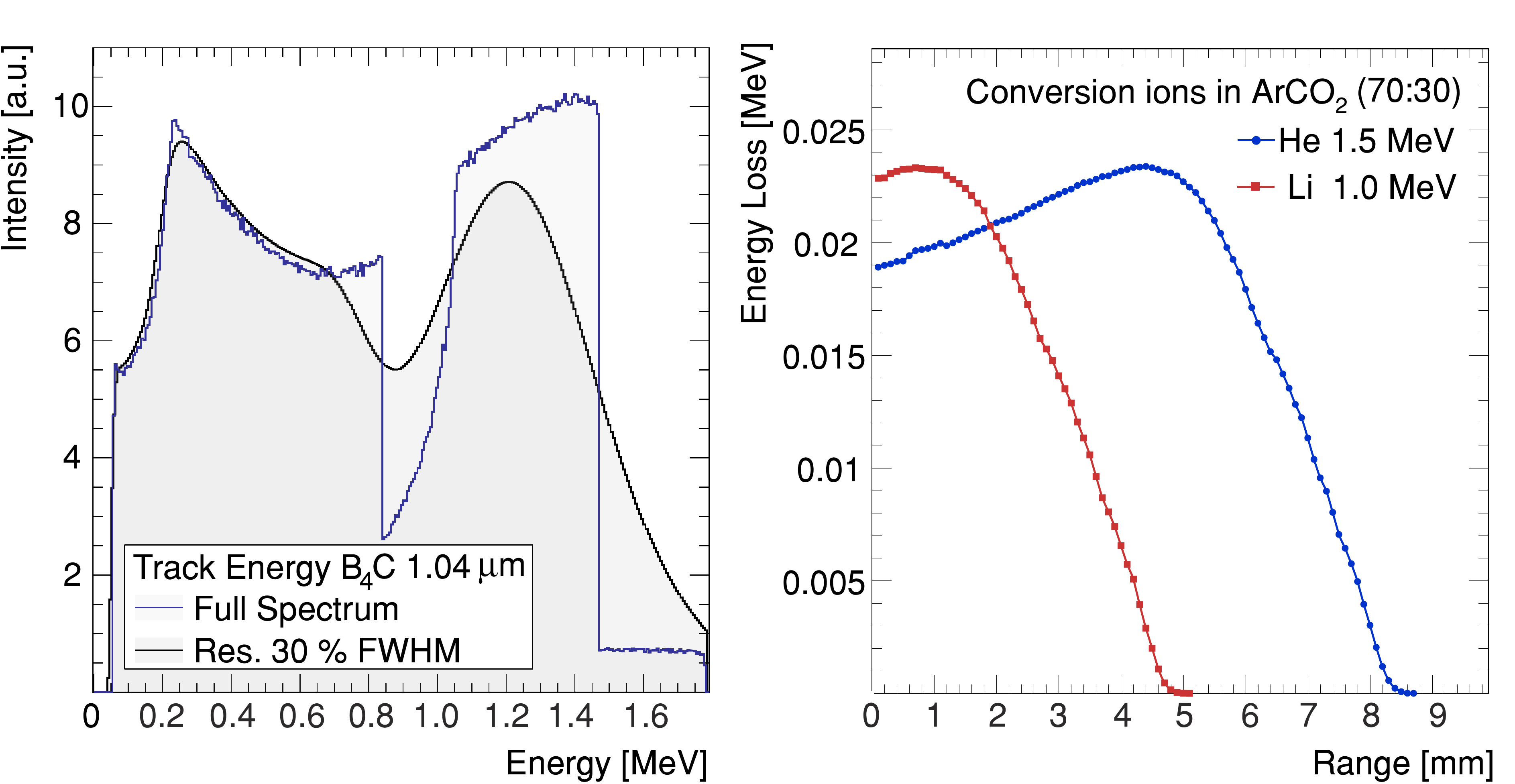}
\caption{Left panel: Energy spectrum of conversion ions after leaving a solid conversion layer of boron (blue and light gray). The peak at higher energies is due to alpha particles, the broader plateau at lower energies is to the largest part comprised of lithium particles and additionally the tail from the peak above. This spectrum is folded by a resolution function (black and dark gray) in order to demonstrate the smearing out of specific features. Right panel: Ionization density in the gas for both conversion ions starting at the maximum energy calculated by SRIM~\cite{SRIM}. Exiting the layer at lower energies corresponds to a translation of the curves to left.}
\label{fig:Ranges}
\end{figure}   
After leaving the solid layer, the ion spectrum is continuous due to the interaction length distribution in the boron, which is a function of conversion depth of the neutron and escape angle of the conversion pair. Due to the lower initial energy and larger loss the lithium ions make up the lower boundary of the energy spectrum. For layer thicknesses beyond 1.2\,$\muup$m a low energy pile-up appears and the further contribution of lithium ions to the detectable signal is reduced. Fig.~\ref{fig:Ranges} (left) shows exemplarily the conversion ion track energy distribution for a layer thickness according to the detector presented here, for which the spectra were calculated by the neutron detector Monte-Carlo transport simulation URANOS~\cite{myself2}. The primary charge carrier generation in the gas then follows the Bethe-Bloch equation. Due to different energy losses for both conversion ions there are qualitatively different range relations and charge distributions. Fig.~\ref{fig:Ranges} (right) shows that the lithium ion starts, even in the case of no energy loss within the boron, already on the descending branch of the Bragg peak, whereas for the alpha particle the ionization density can culminate on its trajectory.

\section{The detector concept}
\subsection{Geometry}
\begin{figure}[ht]
\centering 
\includegraphics[width=\linewidth]{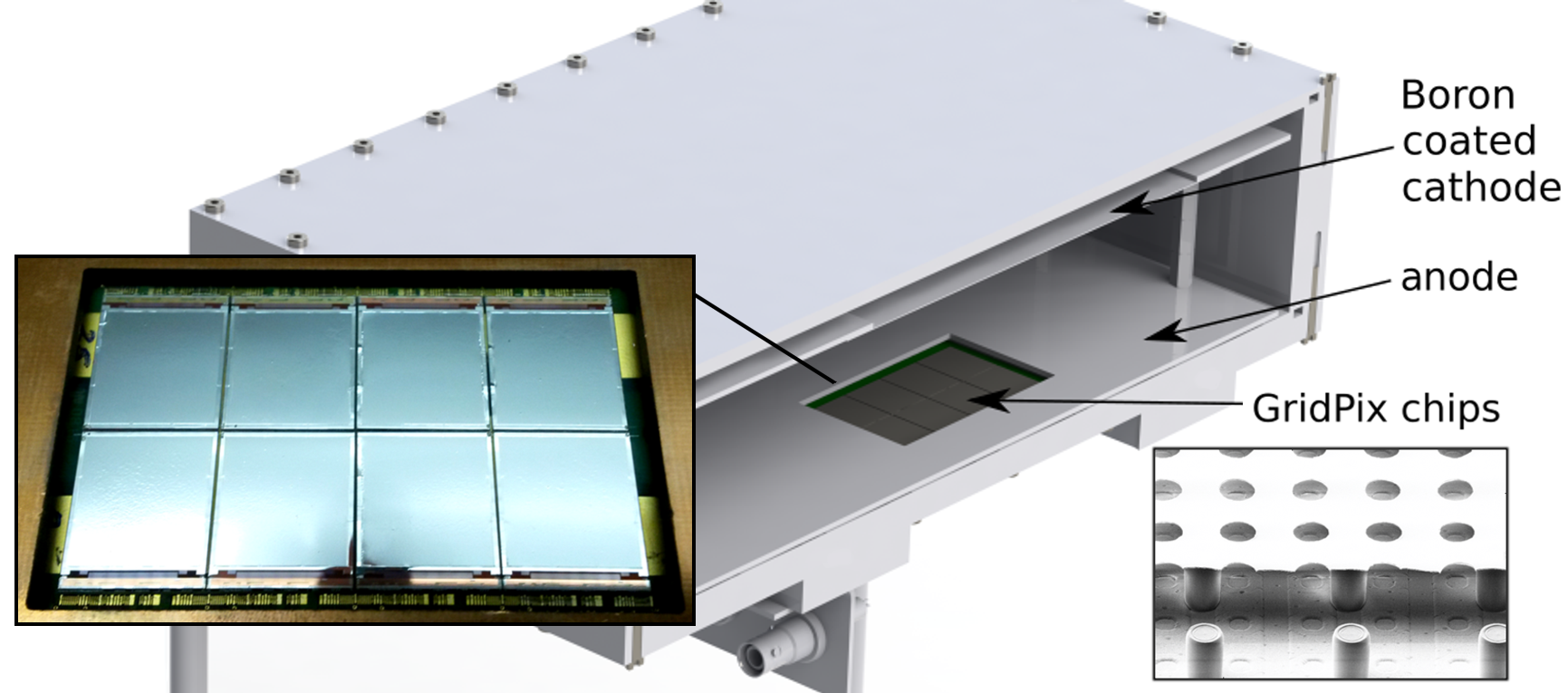}
\caption{Test detector setup: An aluminum sheet coated by $^{10}$B$_4$C acts as a drift cathode. The conversion ions produce tracks of primary ionization in the Ar:CO$_2$ gas, which are then projected onto the anode plate with an inset of the TimePix and InGrid board~\cite{ingridProc}. These chips are closely arranged on the carrier with each of them featuring an individual micromesh gas amplification stage ontop.}
\label{DetectorOpen}
\end{figure}

The detector uses a cathode of 96\,\% enriched $^{10}$B$_4$C in a thickness of 1.04\,$\muup$m aligned in parallel to the readout at a distance of 3.8\,cm. A homogenous drift field of 420\,V/cm projects the conversion products onto the anode plane which is equipped with a highly granular spatial and time resolved readout. The whole assembly, see Fig.~\ref{DetectorOpen}, is placed in an aluminum housing flushed with Ar:CO$_2$ in the mixtures 70:30 and 80:20. The detector was place in front of a moderated AmBe source, which provides a strongly divergent thermal neutron flux with the large epithermal component typical for such setups. 

\subsection{Gas Amplification Stage and Readout}%
Achieving precise spatial and time information on the basis of an event-by-event analysis is improved by one order of magnitude compared to classical pad or wire-based systems by high resolution micropattern gaseous detectors~\cite{BrilliantMicropattern}. We chose to use a derivate of the highly granular TimePix ASIC (application ), which was developed for single photon counting. It features a matrix of $256 \times 256$ pixels with a pitch of 55\,$\muup$m and thus an active area of approximately 2\,cm$^2$. Each of the pixels contains a charge sensitive amplifier and a discriminator plus the entire counting logic needed for time (TOA - time of arrival) or charge measurements (TOT - time over threshold). The principle of the used GridPix~\cite{gridpix1} here is to fabricate a micromesh~\cite{micromegasFirst} as a singular gas amplification stage in a distance of 50\,$\muup$m ontop of a readout structure: the so-called InGrid~\cite{ingridProc2}, a postprocessed aluminum strainer, is manufactured by the IZM Berlin on top of the chip and directly aligned with the charge sensitive input of each TimePix pixel anode of 20\,$\muup$m diameter, see also the lower inlay of Fig.~\ref{DetectorOpen}. A potential difference of 420\,V in Ar:CO$_2$ 80:20 leads to a gas gain of 10$^3$. Each pixel features a dynamic range of 11810 counts, either in TOT, TOA or hit counting mode. The whole matrix is read out as an entire frame, controlled by shutter signal, and the chip clock is operated at 40\,MHz. This results in a time resolution of 25\,ns and a maximum acquisition time of approximately 300\,$\muup$s. Eight TimePix chips are aligned in an array of $2\times 4$ chips, called Octoboard carrier~\cite{pixelTPCIEEE}, with $\approx 460,000$ pixels in total and an active area of approximately 15\,cm$^2$. We chose to operate 5\,\% of the pixels distributed randomly in the TOA mode and the others in the TOT mode. This leads to a pseudo-3D information on each pixel, for which the respectively missing information is extrapolated by its neighbors.

\subsection{Data Acquisition}%
The readout of the TimePix chips, arranged on one Octoboard, is realized by a combination of a computer and an FPGA (\textit{field programmable gate array}) based unit, which acts as a communication interface, first level trigger and data compression architecture. This task is solved by the Scalable Readout System~(SRS)~\cite{SRSOverview}, which is designed to instrument a single system and likewise much larger arrangements of such. 
\begin{figure}[ht]
\centering 
\includegraphics[width=\linewidth]{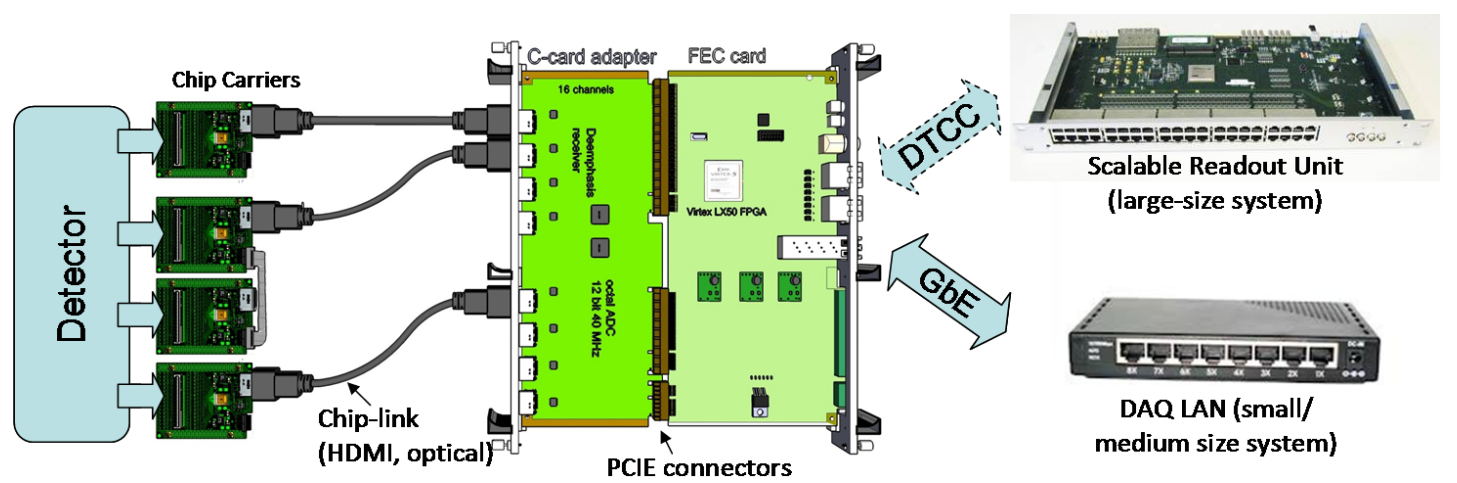}
\caption{The SRS system: Several independent ASICs which make up the detector, are connected via HDMI to the C-card and then to the central data acquisition unit, the FEC card, which hosts a fast FPGA for reading out all subsystems in parallel. The system can either communicate with a computer directly via Ethernet or can be grouped to larger arrays by a Scalable Readout Unit~\cite{SRSOverview}.}
\label{fig:SRSOverview}
\end{figure}
The SRS Hardware is modular, see Fig.~\ref{fig:SRSOverview}. The Octoboard, on which the GridPix chips are mounted, is connected to an intermediate board, which acts as a customizable breakout unit for our subdetectors. Two HDMI connectors are used for communication with the readout system - one for the slow control, the other one for data transfer. The slow control contains all instructions that will be given to the Octoboard. It is also possible to dischain some of the signals, for example the shutter, in order to facilitate the use of an externally generated trigger. The intermediate board is connected to the so-called C-card, serving as a communication interface for HDMI data transfer to the front end card (FEC) for up to four Octoboards or 32~TimePix chips. The front end card features a Virtex-6 FPGA~\cite{virtex6sheet}, which hosts the state machine for the slow control and the data acquisition and zero-suppression unit. The system is run by the TimePix Operation Firmware~(TOF), which communicates with the computer via GBit Ethernet by the TimePix Operation Software~(TOS), both currently under development~\cite{TimePixInSRS} at the Physikalisches Institut Bonn for next-generation Time Projection Chamber instrumentation~\cite{pixelTPC}.
Additionally the system is equipped with an external post mortem trigger, which consists of an amplifier connected to the grid, a discriminator and a microcontroller which closes the shutter after a given time after the first event.

\section{Event Reconstruction}

\begin{figure}[ht]
\centering 
\includegraphics[width=\linewidth]{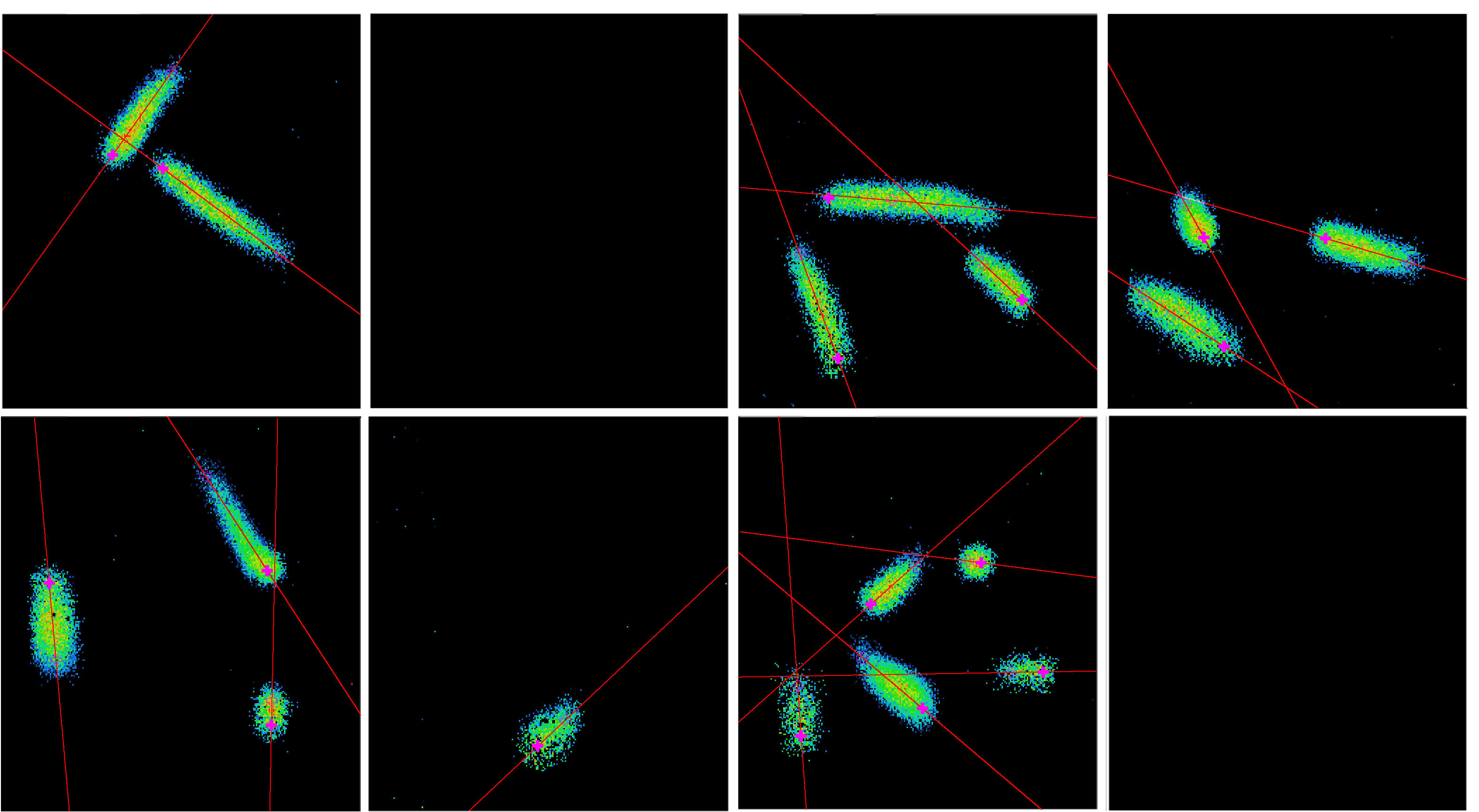}
\caption{The event display showing exemplarily a collection of neutron conversions on the full Octoboard with Ar:CO$_2$ 80:20 at 350\,V grid voltage. Two of the chips are disabled.}
\label{fig:EventDisplay}
\end{figure}

\begin{figure}[ht]
\centering 
\includegraphics[width=0.8\linewidth]{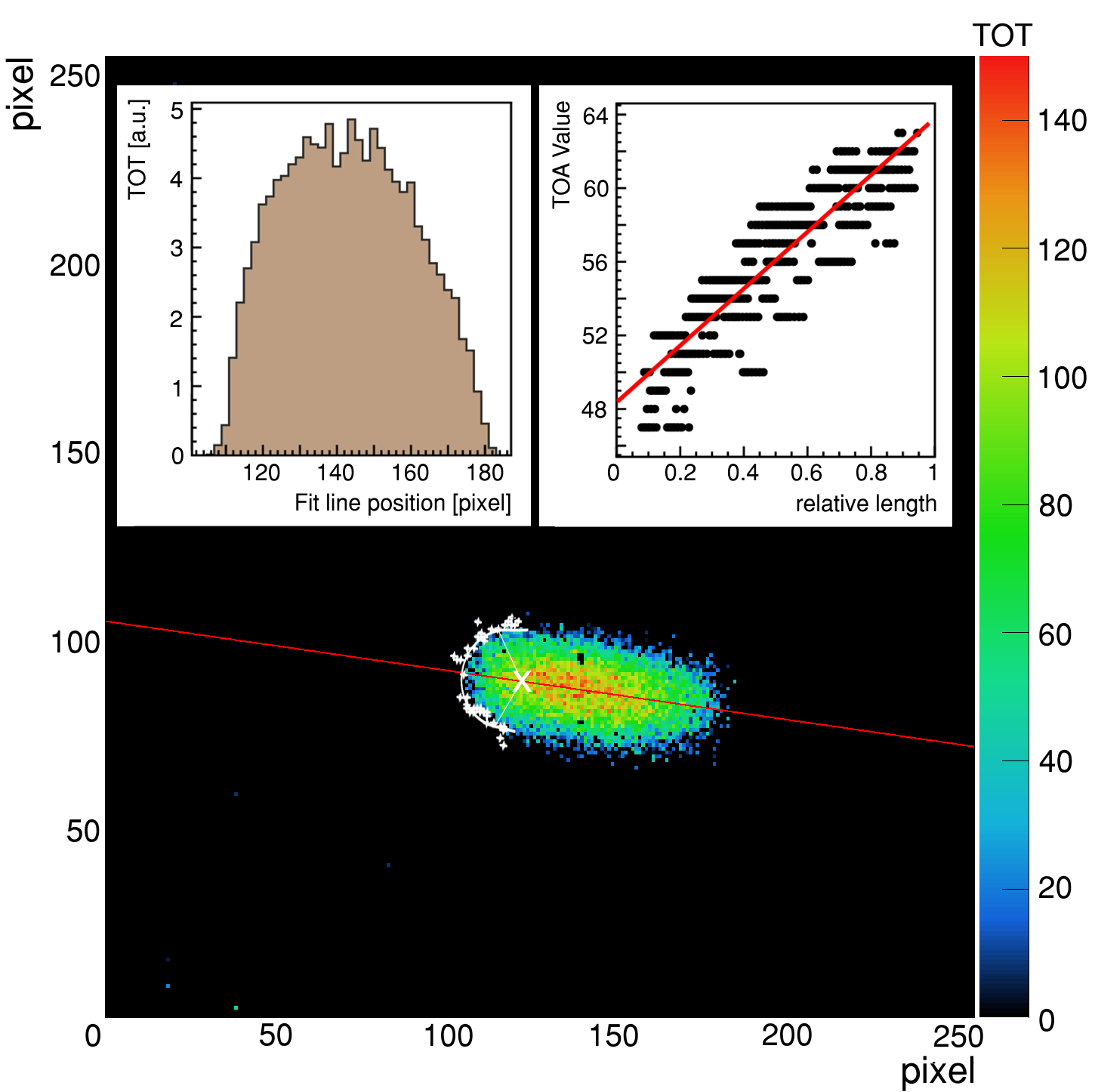}
\caption{Reconstruction of a helium track on a single TimePix chip. A straight line fit through the TOT values (red) allows the projection of the deposited charge (top left panel). By extrapolating the TOA values (top right panel) the drift time can be determined as well as conversion origin. On that side a circular fit (white) to the projected perimeter of the track provides the end of the charge cloud from which the origin is located by an inward shift according to the diameter.}
\label{fig:helium}
\end{figure}

\begin{figure}[ht]
\centering 
\includegraphics[width=0.8\linewidth]{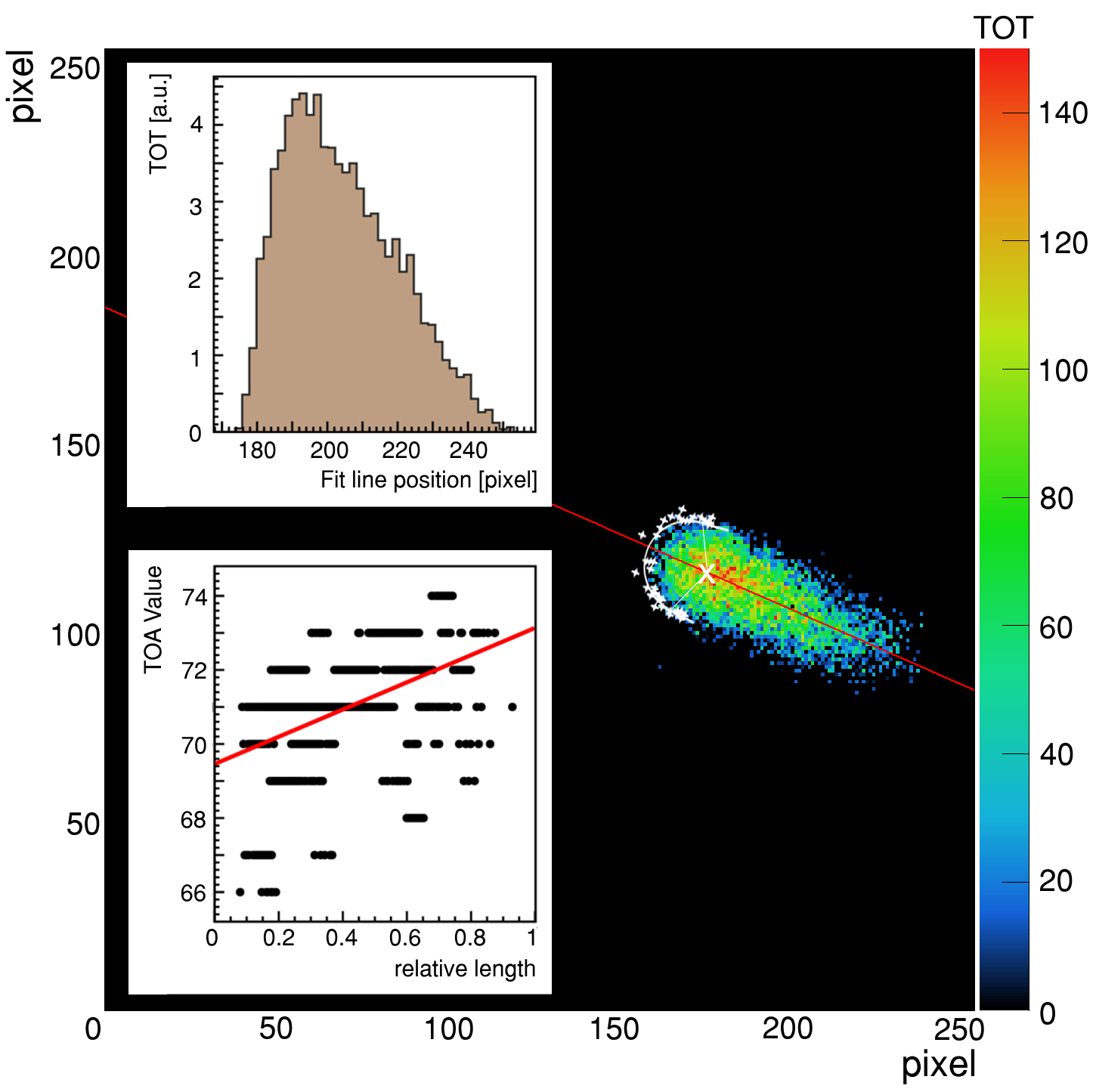}
\caption{Reconstruction of a lithium track on a single TimePix chip. The same methodology is applied as in Fig.~\ref{fig:helium}.}
\label{fig:lithium}
\end{figure}

\begin{figure}[ht]
\centering 
\includegraphics[width=\linewidth]{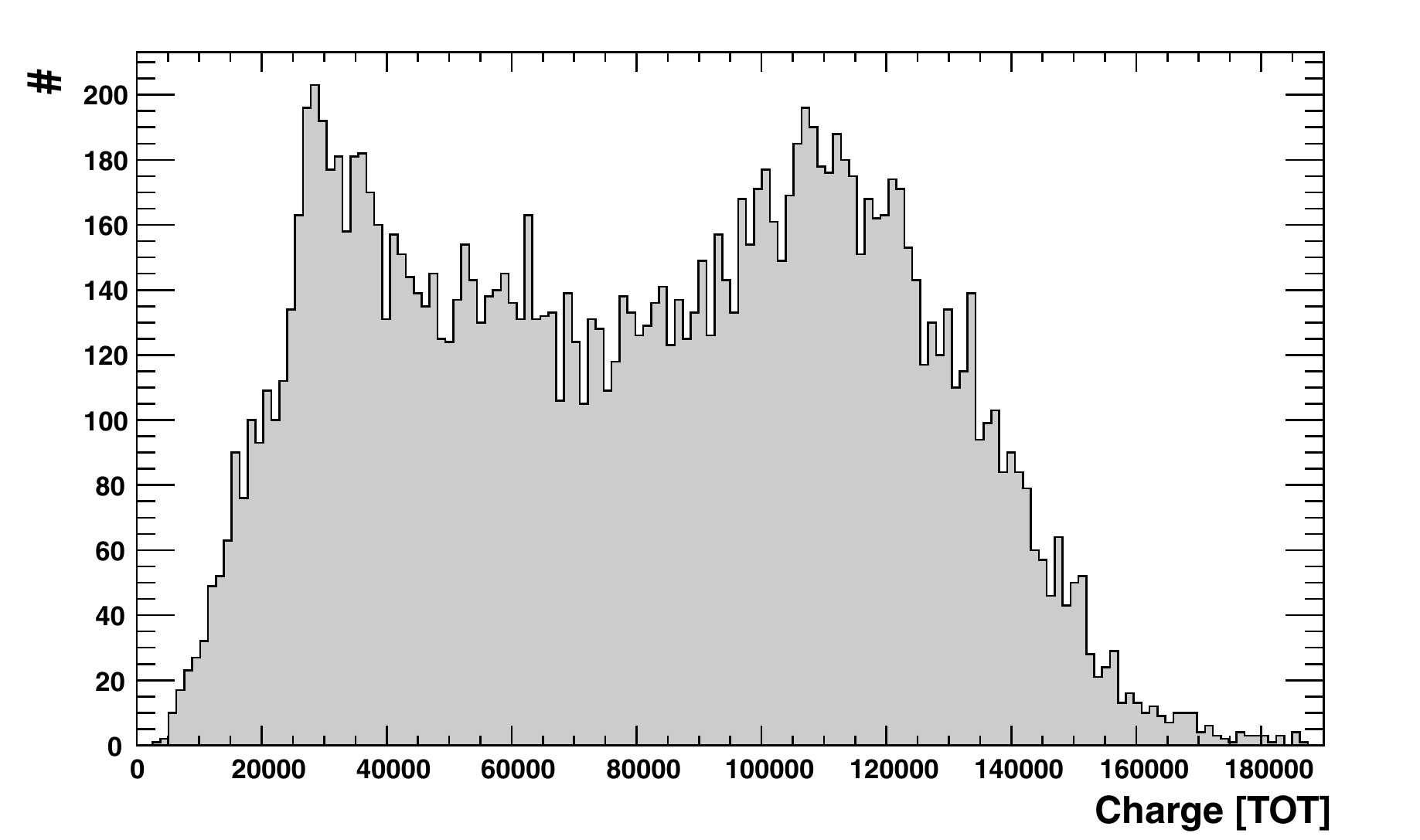}
\caption{Track energy distribution after the event reconstruction showing the full spectrum with the lithium and alpha peak. Compare also to simulations depicted in Fig.~\ref{fig:Ranges}.}
\label{fig:dedx}
\end{figure}

On basis of raw data the event identification is initiated by reconstructing interrelated areas using the DBSCAN~\cite{dbscan} algorithm and the ROOT~\cite{root} framework, both implemented in C++. Such a randomly selected collection of events is shown in Fig.~\ref{fig:EventDisplay}. It has to be noted, that although the later described algorithm intrinsically rejects gamma signatures as they are not 'track-like', the TimePix plus InGrid combination can be operated at voltages which are high enough to detect highly ionizing particles but too low to produce any signal from most of the electrons, muons or photons. These background particles have an energy loss in the detection volume by orders of magnitude lower than such from ions, so the ionization density likewise is much lower. The pixels making up the tracks of Fig.~\ref{fig:EventDisplay} therefore are in fact only firing if a multitude of primary charge carriers arrive at the same time.
As the InGrid technology allows low threshold detection down to a few electrons~\cite{ingridNim}, the projected track image, due to diffusion, consequently resembles an ellipse with loosely connected pixel on its seam. Therefore the event reconstruction has to take into account, that the perimeter of the track is composed of a series of outliers of the bulk. The reconstruction then follows the steps:
\begin{itemize}
	\item Fitting a straight line through the TOT values. If the residuals reveal a slight curvature, the track is cut and if line fits on the segments do still exhibit a curvature, the track is rejected. The latter procedure is a safety measure for field distortions.
	\item Fitting a straight line through the TOA values. Lower values indicate the starting time.
	\item Fitting a circle to the outermost pixels of the side, which was flagged as the origin by the TOA identification. 
	\item Intersecting circle and TOT line and locating the origin by shifting this point inwards along main axis by a distance determined by the track diameter and inclination.
\end{itemize}
For this measurement we only consider non-intersecting, e.g. separable, events. The result of this procedure is presented in Fig.~\ref{fig:helium} and Fig.~\ref{fig:lithium}. Both show in a 2D projection the TOT values of a helium ion and a lithium ion, respectively, with all steps of the event analysis. In the small inlays the energy loss distribution (d$E$/d$x$) along the track is shown by the (brown) histogram and additionally the TOA value graph, which has by configuration only 1/20 of the total information. The curvature in this plot is result of a time-walk effect as already the earliest electron arriving triggers the time counter, so high electron densities appear statistically earlier than areas of lower concentrations. In both track examples the energy loss measurement clearly allows the identification of the particle species, compare also to the right panel of Fig.~\ref{fig:Ranges}. 
In the further analysis it is required for an event to have at least 450 pixels (noise and photon rejection), a non-circular shape (photon and fast neutron recoil rejection on heavier elements like the gas itself), a maximum length of 10\,mm (fast neutron proton recoil rejection due to their CSDA range, see also Fig.~\ref{fig:Ranges}), a minimum mean width of 6 pixels (electron rejection) and a non-injective starting point assignment, e.g. a non-zero TOA slope.
Due to the reasons mentioned above in this chapter the system can be operated at voltages low enough to intrinsically suppress the background, but here the data was taken at a variety of voltages and different gas mixtures in order to ensure the quality of the event reconstruction. This can be demonstrated by the energy loss distribution shown in Fig.~\ref{fig:dedx}, which in comparison with the left panel of Fig.~\ref{fig:Ranges} clearly identifies the measured events as neutron conversion. It has to be noted, that for higher energy alpha particles the probability for the track to be cut at the edge of the sensor increases. Therefore the alpha peak appears to be partly shifted downwards in the energy spectrum. Nevertheless there is no specific cut on the measured event energy, so the lower end of the spectrum is due to the criteria mentioned above.

\section{Spatial Resolution}
\label{SpaRes}

\begin{figure}[ht]
\centering 
\includegraphics[width=\linewidth]{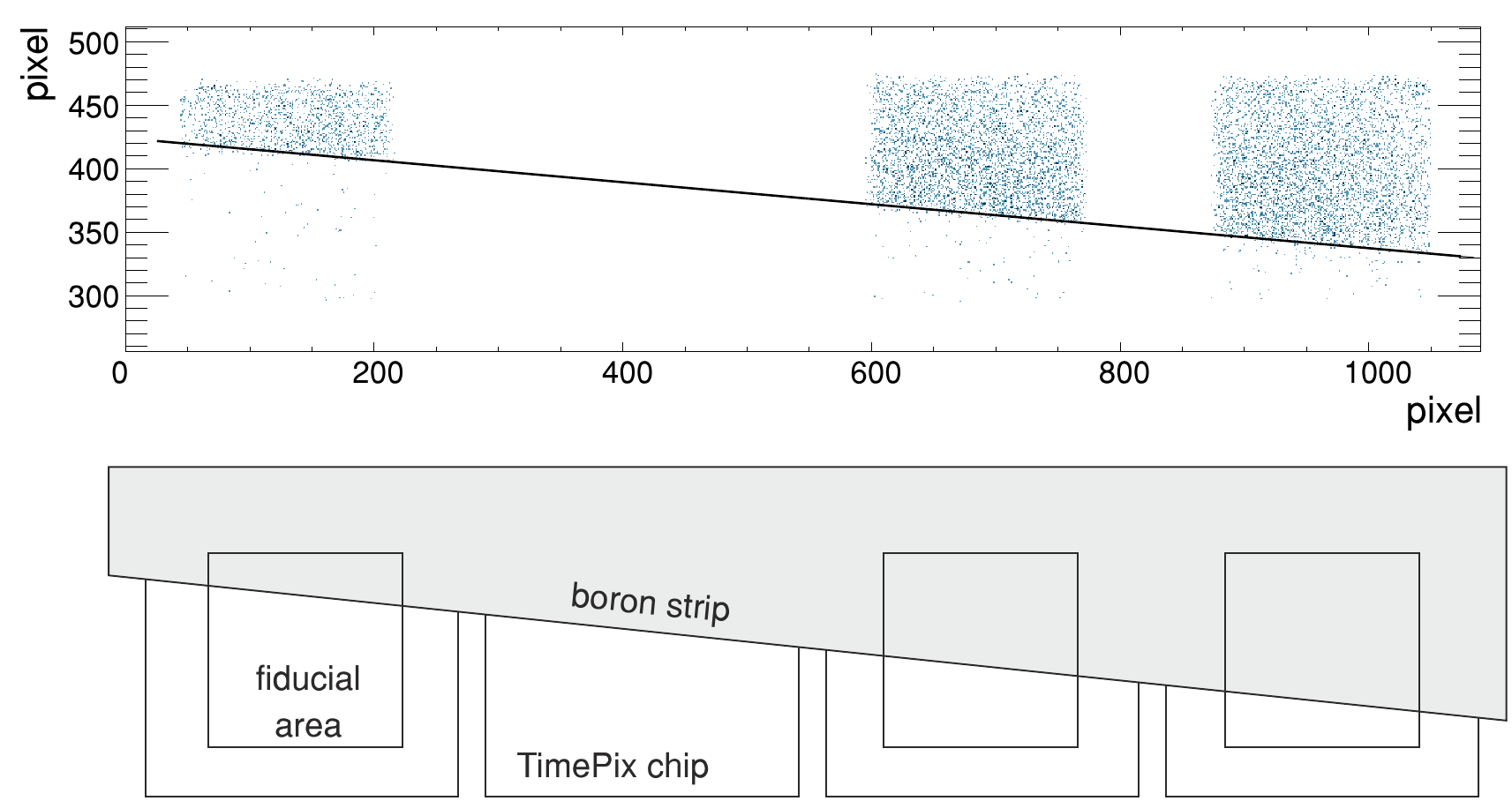}
\caption{Determination of the spatial resolution by a boron coated strip. The lower panel shows in the setup sketch how the edge was aligned over one row of the Octoboard. The fiducial area is about half the size of the entire chip. The reconstructed starting points (blue points) are shown in the upper panel with a line fit to their edge (black). As seen in Fig.~\ref{fig:EventDisplay} one of the chips is disabled.}
\label{fig:StartingPointHisto}
\end{figure}

The spatial distribution was determined by replacing the homogeneous cathode by a boron strip with a clearly defined edge. Fig.~\ref{fig:StartingPointHisto} shows the setup in the lower panel. For the analysis the total area of starting points was fiducialized in order to select tracks which were projected correctly onto the chip. The result is depicted in the upper panel of Fig.~\ref{fig:StartingPointHisto}, which shows a homogeneous distribution of reconstructed neutrons in the boron coated area and a largely empty space below the strip. A line was fitted to the data by calculating the residuals to this function and fitting an error function to the histogram of those whereas the parameters for the best line fit were minimized to yield an abscissa shift for the edge function close to zero and a slope minimizing the variance around the edge. 

\begin{figure}[ht]
\centering 
\includegraphics[width=\linewidth]{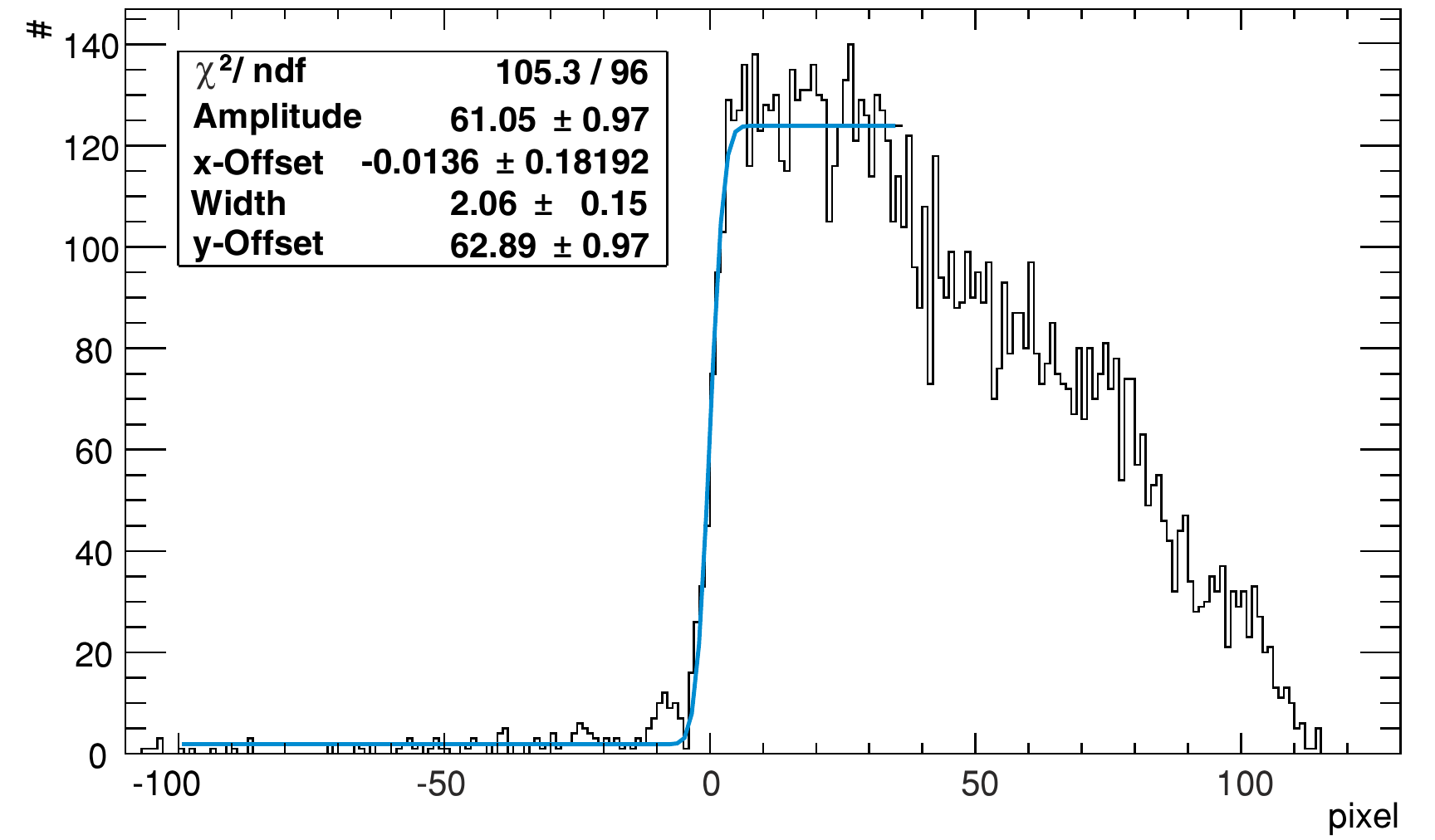}
\caption{Projection of the measured intensity along the line shown in Fig.~\ref{fig:StartingPointHisto} with an error function fitted within the interval of the edge.}
\label{fig:edgeDistance}
\end{figure}

The edge function fitted to the data is defined as the error function:
\begin{equation}
e(k) = \frac{A}{\sqrt{\pi}}\int_{-k}^k e^{-t^2}\text{d}t + c\hspace{0.4cm}\text{with\ } k = \frac{x-x_0}{\sqrt{2}\sigma}
\label{eq:errorFkt}
\end{equation}
with the amplitude $A$ and offset $c$. The parameter $k$ describes the shift on the x-axis $x_0$ in units of the standard deviation $\sigma$. This standard deviation equals the square root of the variance in the Normal distribution
\begin{equation}
g(x) = \frac{1}{\sqrt{2\pi}\sigma} e^{-\frac{(x-x_0)^2}{2\sigma^2}}
\label{eq:gaussFkt}
\end{equation}
and therefore directly provides the spatial resolution. The result is presented in Fig.~\ref{fig:edgeDistance}, which shows the residuals to the line fit. The edge function was fitted from the no-signal region until the end of the signal plateau, which contains the sum of all three active chips.  
Consecutively we obtain a spatial resolution of
\begin{equation}
\sigma = (2.1 \pm 0.15)\,\text{pixels}\hspace{0.4cm}\text{or\hspace{0.2cm}} \sigma = (115\pm8)\,\muup\text{m}.
\label{eq:res}
\end{equation}
It has to be noted, that instead of calculating the end of the track by fitting a circle to the perimeter of the track, one can simply take the last non-zero pixel in the projection along the track and shift this position inwards by the track width. In that case our result for the spatial resolution was approximately only 20\,\% worse, mostly due to the fact that the starting point is still constrained by the line fit through the track.

\section{Summary and Outlook}
In this paper the concept and first measurements of a novel spatially and time resolved neutron detection system have been described. The working principle is based on projecting the conversion products from a solid boron layer onto a highly granular readout chip. We have realized such by the integration of the recently developed GridPix family - TimePix ASICs with an InGrid micromesh on top for gas amplification. The active area consists of an array of $4\times 2$ chips with in total an area of approximately 15\,cm$^2$ in size and $\approx 460,000$~pixels of 55\,$\muup$m pitch. We have successfully implemented a firmware and software readout for the FPGA-based SRS system. The output, zero-suppressed frames providing charge and drift time information, is fed into a raw data analysis, which reconstructs neutron events by their topology and likewise discriminates against background from other particle species. As demonstrated, the conversion track imaging capabilities allow even the distinction between helium and lithium ions by their energy loss in case of sufficiently high energy deposition. This particle identification algorithm is also verified by finally yielding the typical charge deposition spectrum of boron-10. Using boron carbide coated strips, we could determine the spatial resolution to be $\sigma = (115\pm8)\,\muup\text{m}$. A more simple algorithm performed 20\,\% worse.\\
Although the system as presented was developed for conversion ion detection, the high resolution capabilities can be a significant advance for gadolinium-based detectors needing low energy electron identification. The actual work mainly focused on the track topology. In order to realize a neutron detector with a sufficiently high count rate, the step from TimePix to TimePix3~\cite{TimePix3} has to be undertaken, which offers much higher untriggered data taking. Furthermore we are aiming to construct a system which projects the tracks orthogonally onto the readout which then will be placed not in-beam but aside of it.

\section*{Acknowledgments}
MK thanks Ulrich Schmidt, Physikalisches Institut, Heidelberg University, for providing the boron carbide coated strips and Jannis Weimar for providing the hard- and software for the new trigger unit for the detector.
Part of the measurements were carried out at the ELSA facility operated by the Physikalisches Institut, University of Bonn. The project 'Forschung und Entwicklung hochaufl\"osender Neutronendetektoren' was funded by the German Federal Ministry for Research and Education (BMBF), grant identifier: 05K16PD1.

\section*{References}


\end{document}